\documentclass[%
 aip,
rsi,%
 amsmath,amssymb,
 reprint,%
]{revtex4-1}

\usepackage{graphicx}% Include figure files
\usepackage{dcolumn}% Align table columns on decimal point
\usepackage{bm}% bold math
\usepackage{caption}
\usepackage{subcaption}
\usepackage{braket}
\usepackage{color,soul}

\newcommand{\Tabl}[1]{Table\,\ref{#1}}
\newcommand{\Fig}[1]{Figure\,\ref{#1}}
\newcommand{\Eq}[1]{Equation\,\ref{#1}}
\newcommand{\etal}{\,\emph{et al.} }

\begin{document}

\preprint{AIP/123-QED}

\title {Effect of \textbf{Tensile Strain} in GaN Layer on the Band Offsets and 2DEG Density in AlGaN/GaN Heterostructures}% Force line breaks with \\
\author{Mihir Date}
\thanks{Equal contributing authors}

\affiliation{Department of Metallurgy and Materials Science, College of Engineering Pune, Pune,
India}
\author{Sudipta Mukherjee}
\thanks{Equal contributing authors}
\author{Joydeep Ghosh}
\thanks{Equal contributing authors}
\author{Dipankar Saha}
\author{Swaroop Ganguly}
 \email{swaroop.ganguly@gmail.com}
\author{Apurba Laha}
\affiliation{%
Department of Electrical Engineering,
Indian Institute of Technology Bombay, Mumbai, India%\\This line break forced% with \\
}%


\begin{abstract}
%There has been a lack of clarity, regarding the effect of process-induced strain in the underlying GaN layer on Al$_x$Ga$_{1-x}$N/GaN heterostructure properties, which is addressed here. 
We have addressed the existing ambiguity regarding the effect of process-induced \textbf{tensile} strain in the underlying GaN layer on Al$_x$Ga$_{1-x}$N/GaN heterostructure properties. The bandgaps and offsets for Al$_x$Ga$_{1-x}$N on strained GaN are first computed using a cubic interpolation scheme within an empirical tight-binding framework. These are then used to calculate the polarization charge and two-dimensional electron gas density. Our bandstructure calculations show that it is not possible to induce any significant change in band offsets through strain in the GaN layer. The charge-density calculations indicate that such strain can, however, modulate the polarization charge and thereby enhance the 2DEG density at the AlGaN/GaN hetero-interface substantially, by as much as $25\%$ for low Al mole fraction. 
\end{abstract}

%Use showkeys class option if keyword
\maketitle

\section{INTRODUCTION}
In the past few decades, GaN-based heterostructure device technologies have been developed intensively for high-frequency, high-voltage, and optoelectronic device applications. Alloying this semiconductor with AlN produces Al$_x$Ga$_{1-x}$N where $x$ is the Aluminium mole fraction, and the heterostructure between AlGaN and GaN is the key in realizing High Electron Mobility Transistors (HEMTs). In these devices, a two-dimensional electron gas (2DEG) is formed at the hetero-interface due to large polarization fields coupled with the donor-like surface states\cite{Ambacher,Ibbetson}. The growth of these AlGaN/GaN heterostructures on significantly mismatched substrates has always been a coveted goal. For instance, GaN on a Si substrate would be both economical, compared to SiC say,\cite{Dadgar} and would open up the possibility of heterogeneous integration with CMOS electronics. Now, when Si is used as the substrate, there would be lattice mismatch at the GaN and Si interface, introducing a substrate-induced tensile strain in the GaN layer\cite{Azize}; similarly, GaN on the popular sapphire substrate could also have substrate-induced strain. Conventionally, the GaN layer is thick and the strain therein has therefore been assumed to be insignificant or localized. However, even a thick GaN layer may not be free from growth or anneal process-induced strain. Lee\etal\cite{Lee} reported that \textbf{in AlGaN/AlN/GaN heterostructures, the} strain in \textbf{the} GaN \textbf{layer} significantly affects the 2DEG carrier concentration as well as the electron mobility. \textbf{Intuitively, same effect should be present in AlGaN/GaN heterostructures as well.} Recent studies carried out by Kadir\etal\cite{Kadir} have concluded that it is more accurate to assume the Si substrate as the reference material than GaN in order to account for the stresses at the GaN-Si interface. Liu\etal\cite{Liu}, on the other hand, has suggested that referencing Si does not offer any significant advantage. 

Physics-based models have been developed and extensive simulations performed by several groups to predict the 2DEG density as well as the bare surface barrier height in different GaN-based heterostructures \cite{Gordon,Goyal2012,Goyal2013,Ghosh2017}. Indeed, the 2DEG density depends on the band-gap, lattice mismatch, and the surface state parameters. Now, as mentioned above, the GaN substrate has always been considered relaxed. However, in view of the aforementioned experimental works, it now becomes imperative to investigate and clarify how strain in the GaN layer - due to a mismatched substrate or other process-related reasons - impacts the 2DEG concentration in the AlGaN/GaN heterostructure. It is practical to imagine that the GaN layer could be strained due to a combination of mismatched substrate, defects in the heterostructure, and other residual stresses arising from the growth and processing steps. Strain in the GaN is therefore an input parameter in our calculations that represents the cumulative effect of the these factors. We calculate the dependence of strain and Al mole fraction on the band offsets and thereby simulate the 2DEG density. The next section lays out the bandstructure calculation methodology and the resulting bandgap and band offsets. The following one deals with charge-density simulations that take these as inputs to obtain the polarization charge and 2DEG density.

\section{BANDSTRUCTURE CALCULATION}
The starting point for our empirical bandstructure calculation is the nearest-neighbor tight-binding Hamiltonian $H$ with a 4-atom (2 anion, 2 cation), $sp^{3}$ basis for Wurtzite crystals\cite{Kobayashi}. The matrix elements thereof may be represented as: 
\begin{equation}
V^{nb^\prime}_{mb}(\bar{R})\cong\bra{m,b,\bar{R}}H\ket{n,b^\prime,\bar{R}}
\label{eq:1}
\end{equation}
where, $m,n \in \{s,p_x,p_y,p_z\}$ label the localized orbitals, $b,b^\prime \in \{a,c\}$ label the 'anion' and 'cation', and $\bar{R}$ labels the position vector of the central atom in the basal hexagonal plane of the Wurtzite unit cell. In what follows, we will simply denote these tight-binding parameters as $V(\bar{R})$, keeping the basis indices in the subscript and superscript implicit.

%Here, $\ket{\psi_{\alpha}, c}$ and $\ket{\psi_{\beta}, a}$ are the electronic states, expressed as a Bloch sum of atomic orbitals, corresponding to the cation ($c$) and anion ($a$) respectively. The tight-banding parameters for GaN and AlN were obtained from Coughlan\etal\cite{Coughlan}. The tight-binding parameters for Al$_x$Ga$_{1-x}$N were calculated using a cubic interpolation scheme (c.f. \Tabl{t:1}), and the Hamiltonian matrix is constructed\cite{Kobayashi}. 

\textbf{We construct the Hamiltonian matrix following Kobayashi\etal\cite{Kobayashi}, using tight-binding parameters that they showed to be equivalent to those in the Slater-Koster approach\cite{slater}. The values of the parameters for GaN and AlN were obtained from Coughlan\etal\cite{Coughlan}; these parameters were seen to provide the best match to the experimental data, as described below. The parameters $V(\bar{R};x)$ for Al$_x$Ga$_{1-x}$N for a couple of $x$ values were first obtained by linearly mixing those for $x=0$ and $x=1$ as prescribed in G{\"u}rel \textit{et al}\cite{Gurel}. The parameters $V(\bar{R};x)$ for general $x$ were then calculated using a cubic interpolation scheme.} They are shown in \Tabl{t:1} below. The Hamiltonian matrix is then transformed to a $\ket{m,b,\bar{k}}$ basis following Kobayashi\etal\cite{Kobayashi}, and diagonalized to obtain the bandstructure as usual.

%The bandgaps were obtained by subtracting the eigenenergies at global minimum and maximum respectively, at the $\Gamma$-point.

\begin{table} [!ht]
\centering
\begin{ruledtabular} 
\begin{tabular}{ccddd}
T.B. Parameters $V(\bar{R};x)$&$c_0$&\mbox{$c_1$}&\mbox{$c_2$}&\mbox{$c_3$}\\
\hline

$E_{sa}\cong\bra{s,a}H\ket{s,a}$ &-10.615&4.369&-11.975&8.02\\
$E_{sc}\cong\bra{s,c}H\ket{s,c}$&0.912&0.635&1.929&-1.229 \\
$E_{pa}\cong\bra{p_\imath,a}H\ket{p_\imath,a}$&0.818&-0.434&0.862&-0.583\\
$E_{pc}\cong\bra{p_\imath,c}H\ket{p_\imath,c}$&6.678&0.833&10.175&-6.616\\
$U_{ss}\cong\bra{s,a}H\ket{s,c}$&-1.493&0.094&-1.795&1.166\\
$-U_{sz}\cong-\bra{s,a}H\ket{p_z,c}$&1.771&-1.04&1.829&-1.241\\
$U_{zs}\cong\bra{p_z,a}H\ket{s,c}$&3.752&-0.829&4.554&-3.012\\
$U_{zz}\cong\bra{p_z,a}H\ket{p_z,c}$&3.319&-1.02&3.937&-2.62\\
$U_{\jmath\jmath}\cong\bra{p_\jmath,a}H\ket{p_\jmath,c}$&-0.782&0.419&-0.762&0.516\\
\end{tabular}
\end{ruledtabular}
\caption{Cubic interpolation coefficients for tight-binding matrix elements $V(x)$ for Wurtzite Al$_x$Ga$_{1-x}$N: $V(x)=c_0+c_1 x+c_2 x^{2}+c_3 x^{3}$. The matrix elements are represented in the notation of Kobayashi,\cite{Kobayashi} with $\imath \in \{x,y,z\}$, $\jmath \in \{x,y\}$, and the argument $\bar{R}$ dropped for convenience.}
\label{t:1}%a
\end{table}

\begin{figure} [!ht]
\centering
\includegraphics[width=0.38\textwidth]{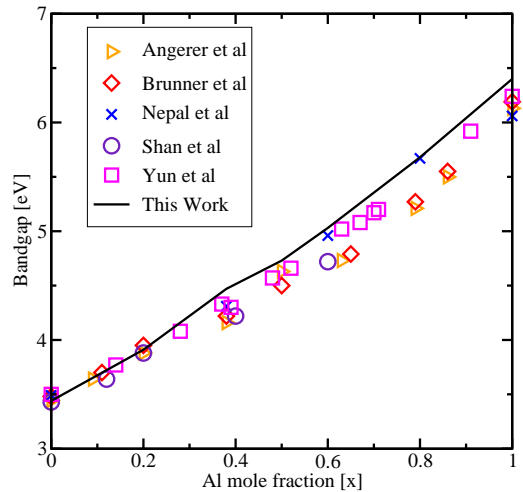}
\caption{Variation in the bandgap of AlGaN with Al mole fraction is shown. Bandgaps obtained from the tight-binding calculations were compared with experiments.\cite{Shan,Bru,Nepal,Angerer,Yun}}
\label{f:1a}
\end{figure}
\begin{figure} [!ht]
\centering
\includegraphics[width=0.38\textwidth]{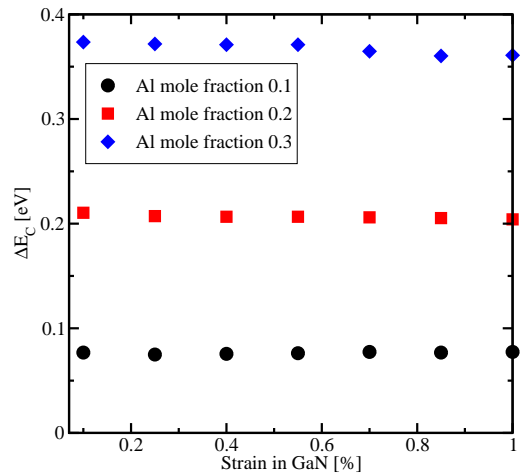}
\caption{The insignificant variation of the band offsets with a rise in GaN substrate strain is shown. The Al mole fraction is used as a parameter.}
\label{f:1}
\end{figure}

%The Hamiltonian in \eqref{eq:1} can also be expressed in terms of the T.B. parameters:
%\begin{equation}
%H_{\alpha\beta}(\textbf{k})= V_{\alpha\beta}\sum_{n}\textbf{u}_{n}(\textbf{k})
%\end{equation}
%where, the functions $u_n(\textbf{k})$ are the once that appear as the periodic Bloch functions, also known as phase-factors. One must note that the phase-factor for the orbitals in the same atom will be zero and hence $H_{\alpha\beta}$=$V_{\alpha\beta}$ for all the diagonal elements appearing in the tight-binding Hamiltonian. Such T.B. terms/parameters are denoted in Table I by $E_{\gamma\delta}$ where $\gamma$($\delta$) specifies the orbital (anion/cation).\\

The bandgaps calculated using the above method for relaxed Al$_x$Ga$_{1-x}$N (\Fig{f:1a}) 
are seen to match quite closely with several reported experimental results. This provides validation for our cubic interpolation methodology.

Now, for strained Al$_x$Ga$_{1-x}$N, the on-site energies may be assumed to remain unchanged since they correspond to the atomic energies, while the the coupling parameters $U$ from \Tabl{t:1} are modified per Harrison's $d^{-2}$ rule\cite{Harrison}:
\begin{equation}
U_{strained}= U_{unstrained} {\left(\frac{d_{0}}{d}\right)}^{2}
\label{eq:2}
\end{equation}
$d$ ($d_0$) being the strained (unstrained) lattice constant.  Our calculations are meant for typical device applications where the Al$_x$Ga$_{1-x}$N film thicknesses would be less than 25nm, implying pseudomorphic Al$_x$Ga$_{1-x}$N\cite{Goyal2013} wherein the strain is given by:
% \begin{equation}
% \varepsilon_{AlGaN}=\dfrac{d(AlGaN)-d_0(AlGaN)}{d_0(AlGaN)}\\
% \label{eq:3}
% \end{equation}
\begin{equation}
\begin{split}
\varepsilon_{AlGaN} 
 &=\dfrac{d(AlGaN)-d_0(AlGaN)}{d_0(AlGaN)} \\
 &=\dfrac{d(GaN)-d_0(AlGaN)}{d_0(AlGaN)}
\end{split}
\label{eq:3}
\end{equation}
\textbf{We point out that uniform scaling of the bond lengths and coupling parameters may be expected to lead to an overestimation of the effect of strain. This is because there is a compression along the c-axis for biaxial tensile strain in the basal hexagonal plane. However, the former is only about a fifth of the latter, due to a Poisson ratio of about 0.2 \cite{Lee}. Further, the results of the following band offset calculations indicate that our assumption - of all the bond lengths scaling like the sides of the hexogonal base - do not make a material difference.}
\textbf{In this work, band offsets have been defined as the difference between the conduction band edges of strained Al$_x$Ga$_{1-x}$N and strained GaN. We have first estimated the bandgaps from the bandstructure. A fraction thereof, viz. 0.63 \cite{Ambacher}, then gives the conduction band offset; this fraction was calibrated to match the experimentally reported 2DEG density \cite{Gordon} using the calculation to be described in the following section. \Fig{f:1} shows that while the band offset varies significantly with the Al mole fraction as expected, it changes only slightly as a function of the strain in the GaN layer. This is because, when the GaN is tensile strained, the conduction band minima of Al$_x$Ga$_{1-x}$N shifts nearly as much as that of the GaN. We note that the relative insensitivity of the band offsets to GaN-layer strain renders our prior overestimation harmless. In the following section, we proceed to perform simulations based on the obtained results in order to estimate the 2DEG density.}

%\begin{table} [!ht]
%\centering
%\caption{The increment of AlGaN band gap with Al mole fraction ($x$).
%Our results are compared with several published results.}
%\begin{ruledtabular}
%\begin{tabular}{ccddddd}
%Al mole fraction& Shan&Brunner&Nepal&Angerer&Yun&This work\\
%$x$&\cite{Shan}&\cite{Bru}&\cite{Nepal}&\cite{Angerer}&\cite{Yun}&This work\\
%\hline
%0 & 3.43 & 3.48 & 3.49 & 3.444 & 3.5 & 3.44 \\
%0.2 & 3.88 & 3.95 & NA & 3.88 & NA & 3.91 \\
%0.38 & 4.22 & 4.22 & 4.31 & 4.16 & 4.32 & 4.47 \\
%0.5 & NA & 4.5 & NA & 4.63 & 4.61 & 4.73 \\
%0.6 & 4.72 & NA & 4.96 & NA & NA & 5.03 \\
%0.8 & NA & 5.27 & 5.67 & 5.21 & NA & 5.68 \\
%1.0 & NA & 6.19 & 6.06 & 6.13 & 6.24 & 6.4 \\
%\end{tabular}
%\end{ruledtabular}
%\label{t:3}%a
%\end{table}

%\begin{figure*} [!ht]
%\includegraphics[width=1.0\textwidth]{FINAL_XC_JOINED_MIHIR.png}
%\caption{The increment of 2DEG density with the strain in the GaN substrate is predicted. The Al mole fraction ($x_c$) is used as a parameter.} 
%\label{f:2}
%\end{figure*}

\begin{figure*}
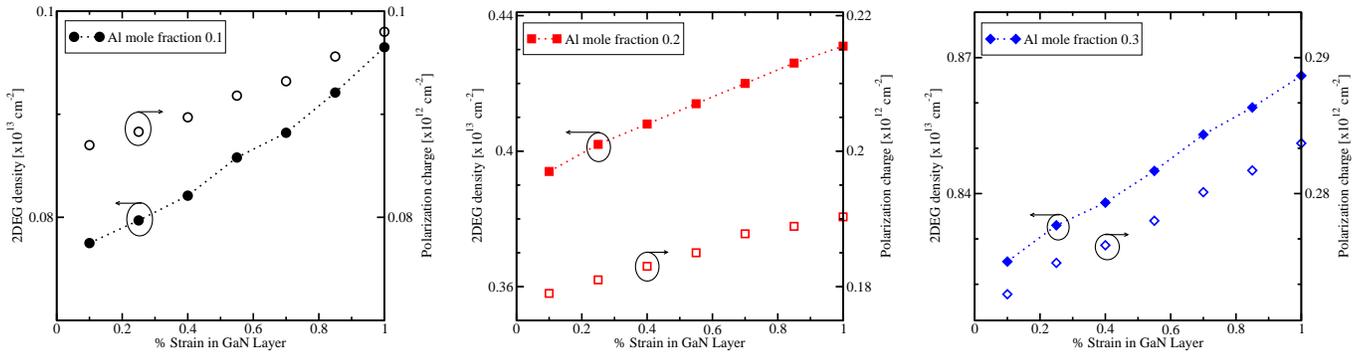

    \centering
    \begin{subfigure}[b]{0.32\textwidth}
        \centering
        \includegraphics[width=\textwidth]{2a_.eps}
        %\caption{$y=x$}
        \label{f:2a}
    \end{subfigure}
    \hfill
    \begin{subfigure}[b]{0.32\textwidth}
        \centering
        \includegraphics[width=\textwidth]{2b_.eps}
        %\caption{$y=3sinx$}
        \label{f:2b}
    \end{subfigure}
    \hfill
    \begin{subfigure}[b]{0.32\textwidth}
        \centering
        \includegraphics[width=\textwidth]{2c_.eps}
        %\caption{$y=5/x$}
        \label{fig:2c}
    \end{subfigure}
    \caption{The increment of 2DEG density and the polarization charge $\sigma$ \textbf{(as given in \Eq{eq:6})} at \textbf{the} AlGaN/GaN hetero-interface with the strain in the GaN substrate is predicted. The Al mole fraction is used as a parameter, AlGaN thickness is 10nm.}
       \label{f:2}
\end{figure*}

\begin{figure} [!ht]
\centering
\includegraphics[width=0.38\textwidth]{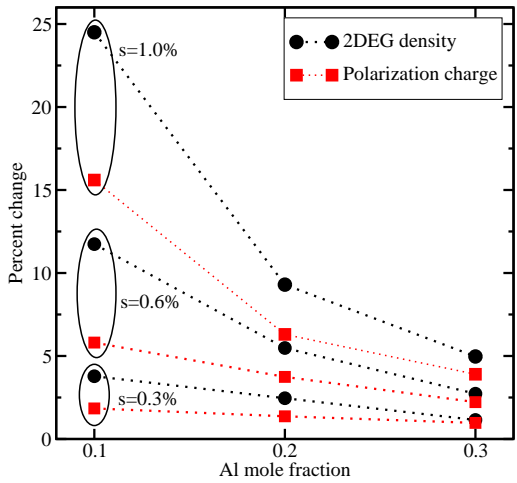}
\caption{\textbf{The percentage increase of the 2DEG density and $\sigma$ compared to the unstrained value is shown. The strain in the GaN layer ($s$) is used as a parameter  (c.f. \Fig{f:2}).}}
\label{f:3}
\end{figure}

%\section{SCHR{\"O}DINGER-POISSON SIMULATIONS}
\section{CHARGE-DENSITY SIMULATION}

In order to study the impact of the GaN substrate strain on the 2DEG density at the Al$_x$Ga$_{1-x}$N/GaN hetero-interface, we make use of the TCAD software package SILVACO\cite{Silv}. Our approach involves calculation of the polarization charge in the heterostructure, followed by self-consistent solution of the one-dimensional Schr{\"o}dinger and Poisson equations along the growth direction. The total polarization charge (\textbf{$\sigma_0$}) at the Al$_x$Ga$_{1-x}$N/GaN hetero-interface is obtained from the expression:
\begin{equation}
% \mathbf{\sigma=|P_{sp0}(AlGaN)-P_{sp0}(GaN)+P_{pe0}(AlGaN)|}
\begin{split}
\mathbf{\sigma_0=|P_{sp0}(AlGaN)-P_{sp0}(GaN)} \\
\mathbf{+P_{pe0}(AlGaN)|}
\end{split}
\label{eq:4}	
\end{equation}
Here, \textbf{$P_{sp0}$ ($P_{pe0}$) denotes} the spontaneous (piezoelectric) polarization charges which \textbf{is} obtained from empirical expressions given by Ambacher\etal\cite{Ambacher}. Thereafter, we follow a procedure due to Gordon\etal\cite{Gordon} to obtain the self-consistent barrier height $\Phi_B$ and 2DEG density $n_s$. Here, the charge density transferred out of the surface donor states is calculated as:
\begin{equation}
n_{surf}=n_0(q\Phi-E_d)
\label{eq:5}
\end{equation}
where $\Phi$ is the estimated surface barrier height, and $n_0$ is the constant surface donor density (cm$^{-2}$eV$^{-1}$) below the surface donor level $E_d$ (eV). Iteratively with this, the SILVACO Schr{\"o}dinger-Poisson solver is used to calculate the 2DEG density n$_{2DEG}$ for a given $\Phi$. The latter is varied until the condition n$_{2DEG}=n_{surf}$ is satisfied. At this point, $\Phi=\Phi_B$ and n$_{2DEG}=n_s$. Note that we have considered a uniform distribution of the surface donor states \cite{Gordon,Goyal2012}. Our results are observed to match well with the experimental data as given in Gordon\etal\cite{Gordon}. 

In the next step, we introduce strain in the GaN layer. This will introduce an additional piezoelectric polarization charge $P_{pe}(GaN)$ in the GaN layer itself. Further, it necessitates a recalculatation of the \textbf{piezoelectric polarization charge term in AlGaN layer. The total polarization charge is thus modified as:} %term \textbf{$P_{pe0}(AlGaN)$} in \Eq{eq:4} \textbf{as:} 
%We assume that it does not alter the AlGaN surface state parameters. 
% \hl{$\sigma=P_{sp}(AlGaN)-P_{sp}(GaN)+P_{pe}(AlGaN)-P_{pe}(GaN)$}.
\begin{equation}
\begin{split}
\mathbf{\sigma=|P_{sp0}(AlGaN)-P_{sp0}(GaN)+P_{pe}(AlGaN)} \\
\mathbf{-P_{pe}(GaN)|}
\end{split}
\label{eq:6}
\end{equation}
\textbf{Where} 

\begin{equation}
\begin{split}
\mathbf{P_{pe}(AlGaN)=2\bigg(\frac{a^{\prime}-a(AlGaN)}{a(AlGaN)}\bigg)\cdot} \\
\mathbf{\bigg(e_{31}(AlGaN)-e_{33}(AlGaN)\frac{C_{13}(AlGaN)}{C_{33}(AlGaN)}\bigg)}
\end{split}
\label{eq:7}
\end{equation} 

\begin{equation}
\begin{split}
\mathbf{P_{pe}(GaN)=2\bigg(\frac{a^{\prime}-a(GaN)}{a(GaN)}\bigg)\cdot} \\
\mathbf{\bigg(e_{31}(GaN)-e_{33}(GaN)\frac{C_{13}(GaN)}{C_{33}(GaN)}\bigg)}
\end{split}
\label{eq:8}
\end{equation}

\textbf{Here, $a$ denotes the bulk lattice constant, 
% the index $l$ signifies the material AlGaN or GaN, 
$C_{13}$ ($C_{33}$) is the elastic constant, and $e_{31}$ ($e_{33}$) is the piezoelectric constant. 
The strained GaN lattice parameter $a^{\prime}$ is: $a^{\prime}=a(GaN)+s\cdot a(GaN)$ with $s$ as the process-induced strain in the GaN layer.} 
Indeed, the strain in the underlying GaN layer would not influence AlGaN surface state parameters. 
% We have calculated the polarization charges at each interface and set it in the simulator. 
% As there was no way of imposing any external mechanical strain to the GaN layer beneath AlGaN, we have taken the best use of model development technique. 
The basic bandstructure parameters for the strained GaN (namely electron affinity and bandgap) and its band offsets with the pseudomorphic Al$_x$Ga$_{1-x}$N were taken from the bandstructure calculation described earlier and provided as inputs to the simulator through its C-Interpreter interface. Thereby, we can obtain the 2DEG density for different Aluminum mole fraction and GaN layer strain.

% \begin{table} [!ht]
% \centering
% \caption{The percentage increase of 2DEG density and polarization $\sigma$ is shown when strain in GaN layer is 1\%, compared to the unstrained value (c.f. \Fig{f:2}).}
% \begin{ruledtabular} 
% \begin{tabular}{ccd}
% \mbox{$x$}&\mbox{\% Change in 2DEG density}&\mbox{\% Change in polarization}\\
% \hline
% 0.1&24.5&15.6\\
% 0.2&9.3&6.3\\
% 0.3&4.97&3.9\\
% \end{tabular}
% \end{ruledtabular}
% \label{t:2}%a
% \end{table}

\Fig{f:2} shows the simulated 2DEG density as a function of strain in the GaN substrate. The Al mole fraction $x$ is used as a parameter, and the AlGaN barrier thickness is set to 10nm. We find that the 2DEG density increases modestly with the strain in the GaN layer for a given $x$. 
\textbf{By following \Eq{eq:6} to \Eq{eq:8}, we} hypothesize that the strain in the GaN layer modulates the strain in the AlGaN layer, and thence the piezoelectric polarization therein. 
\Fig{f:2} also shows the variation of polarization charge $\sigma$ with strain. We find that the increase in $\sigma$ in the AlGaN/GaN hetero-interface more or less tracks that of the 2DEG density, and thereby, explains its origin. 

The percentage increase in the 2DEG density due to strain in GaN layer as a function of $x$ is shown in % \Tabl{t:2}. 
\Fig{f:3}. 
\textbf{We have chosen three values of $s$ to show the impact of GaN layer strain on the 2DEG density.}
% The impact of strain in the GaN layer 
\textbf{This impact}
is seen to decrease sharply with increasing $x$. This happens because when $x$ increases, $\sigma$ increases significantly: both the terms \textbf{$P_{sp0}(AlGaN)$} and \textbf{$P_{pe}(AlGaN)$} are enhanced (c.f. \textbf{\Eq{eq:6}}). Therefore, the impact of the incremental polarization charge due to strain in the GaN layer on $\sigma$ will be marginalized. Our calculations predict maximal increase in the 2DEG at small $x$, as much as 25\% increase in the 2DEG density for $x$=0.1.

\section{CONCLUSION}
The properties of Al$_x$Ga$_{1-x}$N/GaN heterostructures have been computed for the case where the underlying GaN is tensile strained. Bandgaps and band offsets are calculated using the empirical $sp^{3}$ tight-binding method, the polarization charge is calculated, and then a self-consistent Schr{\"o}dinger-Poisson solver yields the 2DEG concentration at the hetero-interface. The strain in the GaN induces incremental strain in the AlGaN layer; this in turn leads to increased piezoelectric polarization. It is found that the band offsets have weak dependence on the GaN layer strain, because the latter induces roughly the same amounts of shift in the GaN and AlGaN conduction band edges. On the other hand, substantial enhancement in the 2DEG density is possible when the GaN layer is strained - this is due to the consequent increase in polarization. For larger Al mole fraction, the polarization is large anyway and the relative boost due to the GaN layer strain becomes smaller; therefore, the enhancement falls steeply with increasing Al mole fraction. Our work suggests that strain in the GaN layer - which might be unintentional today - could be used to achieve higher 2DEG density in the future if it can be engineered controllably. Lastly, it provides a framework to extend this effect to other III-N heterstructures.

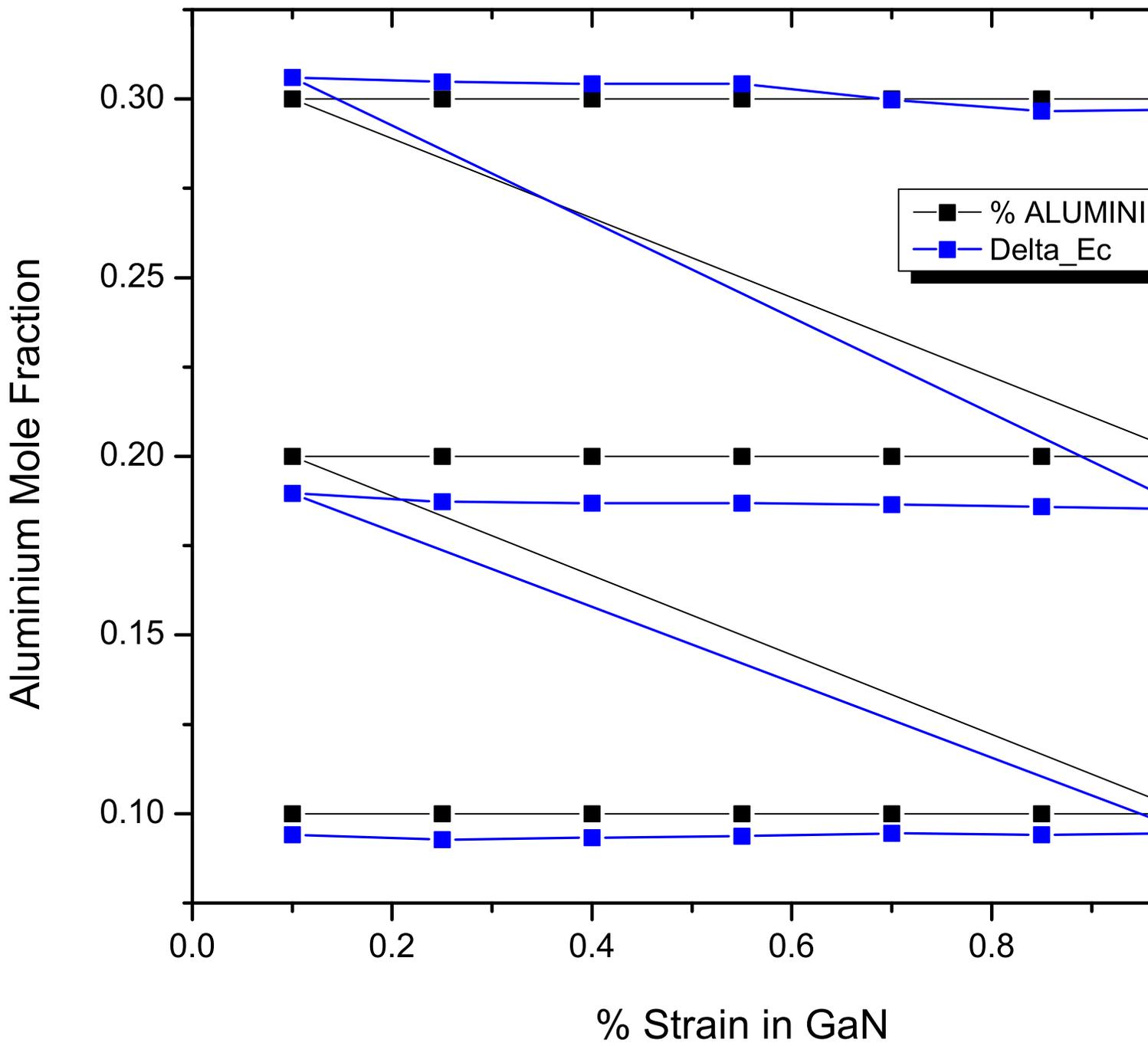
\end{document}